\documentclass[sigconf]{acmart}
\AtBeginDocument{%
  \providecommand\BibTeX{{%
    \normalfont B\kern-0.5em{\scshape i\kern-0.25em b}\kern-0.8em\TeX}}}
\setcopyright{acmcopyright}
\copyrightyear{2021}
\acmYear{2021}
\acmDOI{10.1145/1122445.1122456}

\acmConference[SIGSpatial '21]{SIGSpatial '21: ACM Advances in Geographic Information Systems}{Nov 02--05, 2021}{Beijing, China}
\acmBooktitle{SIGSpatial '21: ACM Advances in Geographic Information Systems}
\acmPrice{15.00}
\acmISBN{978-1-4503-XXXX-X/18/06}

\usepackage{setspace}
\usepackage{placeins}
\usepackage{afterpage}
\usepackage[utf8]{inputenc}
\usepackage{graphicx}
\usepackage{balance}
\usepackage{amsthm}
\usepackage{color}
\usepackage{caption}
\usepackage{amsmath}
\usepackage{tabu}
\usepackage{array}
\usepackage{booktabs}
\usepackage{threeparttable}
\usepackage{relsize}
\usepackage{physics}

\usepackage{algorithm}
\usepackage[noend]{algpseudocode}

\newlength{\maxwidth}
\newcommand{\algalign}[2]
{\makebox[\maxwidth][r]{$#1{}$}${}#2$}

\usepackage{float}
\usepackage{stfloats}
\usepackage{lipsum}
\usepackage[english]{babel}

\usepackage{comment}

\usepackage[T1]{fontenc}
\usepackage{mwe}    
\usepackage{subfig}

\let\oldnl\nl
\newcommand{\nonl}{\renewcommand{\nl}{\let\nl\oldnl}}
\newtheorem{thm}{Theorem}
\newtheorem{problem}{Problem}

\theoremstyle{definition}
\newtheorem{defn}{Definition}

\def\Equal{\texttt{=}}
\DeclareMathOperator{\EX}{\mathbb{E}}

\begin{document}

\title{HTF: Homogeneous Tree Framework for Differentially-Private Release of Location Data}

\author{Sina Shaham$^1$, Gabriel Ghinita$^2$, Ritesh Ahuja$^1$, John Krumm$^3$, Cyrus Shahabi$^1$}
\email{{sshaham,rahuja,shahabi}@usc.edu,gabriel.ghinita@umb.edu,jckrumm@microsoft.com}
\affiliation{%
  \institution{$^1$University of Southern California, $^2$University of Massachusetts Boston, $^3$Microsoft Research}
}

\renewcommand{\shortauthors}{Shaham et al.}
\begin{abstract}
Mobile apps that use location data are pervasive, spanning domains such as transportation, urban planning and healthcare.
Important use cases for location data rely on statistical queries, e.g., identifying hotspots where users work and travel. Such queries can be answered efficiently by building histograms. However, precise histograms can expose sensitive details about individual users. Differential privacy (DP) is a mature and widely-adopted protection model, but most approaches for DP-compliant histograms work in a  data-independent fashion, leading to poor accuracy. The few proposed data-dependent techniques attempt to adjust histogram partitions based on dataset characteristics, but they do not perform well due to the addition of noise required to achieve DP.
We identify {\em density homogeneity} as a main factor driving the accuracy of DP-compliant histograms, and we build a data structure that splits the space such that data density is homogeneous within each resulting partition. We show through extensive experiments on large-scale real-world data that the proposed approach achieves superior accuracy compared to existing approaches.
\end{abstract}


\maketitle
\section{Introduction}\label{sec:intro}

Statistical analysis of location data, typically collected by mobile apps, helps researchers and practitioners understand patterns within the data, which in turn can be used in various domains such as transportation, urban planning and public health. At the same time, 
significant privacy concerns arise when locations are directly accessed. Sensitive details about individuals, such as political or religious affiliations, alternative lifestyle habits, etc., can be derived from users' whereabouts. Therefore, it is essential to account for user privacy and protect location data.

{\em Differential privacy (DP)}~\cite{dwork2006calibrating} is a well-established protection model for statistical data processing. DP allows answering aggregate queries (e.g., count, sum) while hiding the presence of any specific individual within the data.
In other words, the query results do not permit an adversary to infer with significant probability whether a certain individual's record is present in the dataset or not. DP achieves protection by injecting random noise in the query results according to well-established rules. It is a powerful semantic model adopted by both government entities (e.g., Census Bureau) as well as major industry players.

In the location domain, existing DP-based approaches build a spatial index structure, and perturb index node counts using random noise. Subsequently, queries are answered based on the noisy node counts. Building DP-compliant index structures has several benefits: first, querying indexes is a natural approach for most existing spatial processing techniques; second, using an index helps quantify and limit the amount of disclosure, which becomes infeasible if one allows arbitrary queries on top of the exact data; third, query efficiency is improved. 
Due to large amounts of background knowledge data available to adversaries (e.g., public maps, satellite imagery), information leakage may occur both from query answers, as well as from the properties of the indexing structure itself. To deal with the structure leakage, initial approaches used {\em data-independent} index structures, such as quad-trees, binary space partitioning trees, or uniform grids (UG). 
No structural leakage occurred, and the protection techniques focused on improving the signal-to-noise ratio in query answers. However, such techniques perform poorly when the data distribution is skewed.

More recently, data-dependent approaches emerged, such as adaptive grids (AG), or kd-tree based approaches~\cite{AG, inan2010private}. AG overcomes the rigidity of UG by providing a two-level grid, where the first level has fixed granularity, and the second uses a granularity derived from the coarse results obtained from the first level. While it achieves improvement, it is still a rather blunt tool to account for high variability in dataset density, which is quite typical of real-life datasets. Other more sophisticated approaches tried to build kd-trees or R-trees in a DP-compliant way, but to do so they used DP mechanisms such as the exponential mechanism (EM) (discussed in Section 2) which are difficult to tune and may introduce significant errors in the data. In fact, the work in~\cite{AG} shows that data-dependent structures based on EM fail to outperform AG.

Our proposed {\em Homogeneous Tree Framework} (HTF) is addressing the problem of DP-compliant location protection using a data-dependent approach that focuses on building index structures with {\em homogeneous intra-node density}. Our key observation is that density homogeneity is the main factor influencing the signal-to-noise ratio for DP-compliant spatial queries (we discuss this aspect in detail in Section 3). Rather than using complex mechanisms like EM which have high sensitivity, we derive theoretical results that can directly link index structure construction with intra-node data density based on the lower-sensitivity Laplace mechanism (introduced in Section 2). This novel approach allows us to build effective index structures capable of delivering low query error without excessive consumption of privacy budget. 
HTF is custom-tailored for capturing areas of homogeneous density in the dataset, which leads to significant accuracy gains. 
Our specific contributions are:
\begin{itemize}
    \item We identify data homogeneity as the main factor influencing query accuracy in DP-compliant spatial data structures;
    \item We propose a custom technique for homogeneity-driven DP-compliant space partitioning based on the Laplace mechanism, and we perform an in-depth analysis of its sensitivity;
    \item We derive effective DP budget allocation strategies to balance the noise added during the building of the structure with that used for releasing query answers;
    \item We propose a set of heuristics to automatically tune data structure parameters based on data properties, with the objective of minimizing overall error in query answering;
    \item We perform an extensive empirical evaluation showing that HTF outperforms existing state-of-the-art on real and synthetic datasets under a broad range of privacy requirements.
\end{itemize}

The rest of the paper is organized as follows: Section~2 presents background information and introduces the problem definition. Section~3 provides the overview of the proposed framework, followed by technical details in Section 4. We evaluate our approach empirically against the state-of-the-art in Section~5. We survey related work in Section~6 and conclude in Section~7.

\begin{figure}[t]
\raggedright
\includegraphics[scale=.65]{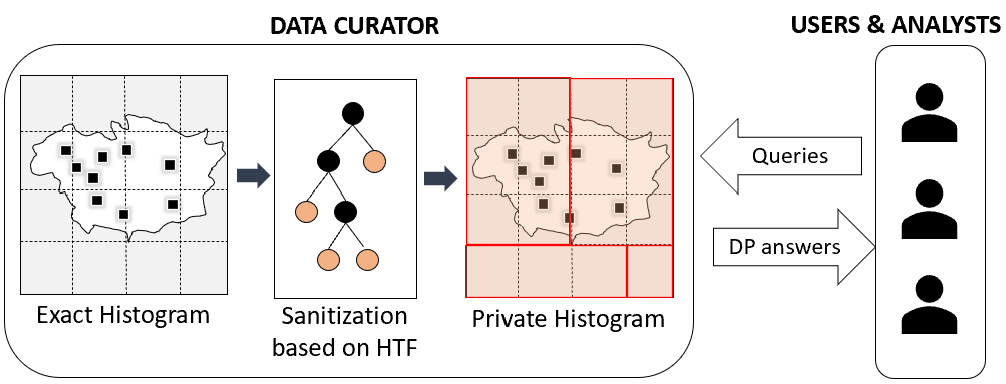}
\centering
\caption{System model for private location histograms.}
\label{Fig: system model}
\vspace{-20pt}
\end{figure}

\section{Background and Definitions}

Private publication of location histograms follows the two-party system model shown in Fig.~\ref{Fig: system model}. The data owner/curator first builds an exact histogram with the distribution of locations on the map. Non-trusted users/analysts are interested in learning the population distribution over different areas, and perform statistical queries. The goal of the curator is to publish the location histogram without the privacy of any individual being compromised. To this end, the exact histogram undergoes a sanitizing process according to DP to generate a {\em private histogram}. In our proposed method, a tree-based algorithm is applied for protection, and the tree's nodes, representing a private histogram of the map, are released to the public.  
Analysts/researchers ask unlimited {\em count queries} that are answered from the private histogram. Furthermore, they may download the whole private histogram, and the protection method remains strong enough to protect the identity of individuals in the database. 

\subsection{Differential Privacy}\label{Sec: differential privacy}


Consider two databases $\mathcal{D}$ and $\mathcal{D}'$ that differ in a single record $t$, i.e., $\mathcal{D}'=\mathcal{D}\bigcup \{t\}$ or $\mathcal{D}'=\mathcal{D}\textbackslash\{t\}$. $D$ and $D'$ are commonly referred to as {\em neighboring} or {\em sibling}. 

\begin{defn}[$\epsilon$-Differential Privacy\cite{dwork2008differential}] 
    A randomized mechanism $\mathcal{A}$ provides $\epsilon$-DP if for any pair of neighbor datasets $D$ and $D'$, and any $S\in Range(\mathcal{A})$,
    \begin{equation}
        \dfrac{Pr(\mathcal{A}(\mathcal{D})=S)}{Pr(\mathcal{A}(\mathcal{D}')=S)} \leq e^\epsilon 
    \end{equation}
    
\end{defn}

Parameter $\epsilon$ is referred to as privacy budget.
$\epsilon$-DP requires that the output $S$ obtained by executing mechanism $\mathcal{A}$ does not significantly change by adding or removing one record in the database. Thus, an adversary is not able to infer with significant probability whether an individual's record was included or not in the database. An important property of DP is {\em composability}~\cite{dwork2014algorithmic}: running in succession multiple mechanisms that satisfy DP with privacy budgets $\epsilon_1, \epsilon_2,...,\epsilon_n$, results in $\epsilon$-differential privacy where $\epsilon = \sum_{i=1}^n \epsilon_i$.

There are two common methods to achieve differential privacy: the {\em Laplace mechanism} and the {\em exponential mechanism (EM)}. Both approaches are closely related to the concept of {\em sensitivity}, which captures the maximal difference achieved in the output by adding or removing a single record from the database.

\begin{defn}[$L_1$-Sensitivity\cite{dwork2006calibrating}]
Given sibling datasets $\mathcal{D}$, $\mathcal{D}'$ the $L_1$-sensitivity of a set $f = \{f_1, \ldots, f_m\}$ of real-valued functions is:

$$\Delta f\Equal \underset{\forall \mathcal{D},\mathcal{D}'}{max}\sum_{i\Equal1}^m |f_i(\mathcal{D})- f_i(\mathcal{D}')|$$
\end{defn}

\newcommand{\rvec}{\mathrm {\mathbf {r}}} 
\begingroup
\begin{table}
\caption {Summary of notations.} 
\vspace{-10pt}
\centering
\begin{tabular}{>{\arraybackslash}m{2cm} >{\arraybackslash}m{5.8cm} }
\hline\hline
  Symbol  & Description\\    \hline
  $\epsilon_{\text{tot}}$ & Total privacy budget\\
  $\epsilon_{\text{height}}, \epsilon_{\text{data}}$ & Height estimation, data perturbation budget\\
  $\epsilon_{\text{prt}},\epsilon_{\text{prt}}',\epsilon_{\text{prt}}''$ & Partitioning budget: total, per level, per round\\ 
  $o_k$ & Objective function output for index k\\
  $c_{ij}$ & Number of data points in row~$i$ and column~$j$\\
  $h$ & Tree height\\
\hline\hline
\end{tabular}
\label{tab:table1}
\vspace{-15pt}
\end{table}
\endgroup

\subsubsection{Laplace Mechanism}

The Laplace mechanism is a widely used technique to achieve $\epsilon$-DP. It adds to the output of a query function $f$ noise drawn from Laplace distribution Lap$(b)$ with scale $b$, where $b$ depends on two factors: sensitivity and privacy budget.

\begin{equation}
    \text{Lap}(x|b) = \dfrac{1}{2b}e^{|x|/b}\; \text{where }\; b=\dfrac{\Delta {f}}{\epsilon} 
\end{equation}
To simplify notation, we denote Laplace noise by $\text{Lap}( \dfrac {\Delta {f} } {\epsilon} )$, as it only depends on the sensitivity and budget.

\subsubsection{Exponential Mechanism}

The exponential mechanism (EM) is another approach to achieve $\epsilon$-DP when the output of a computation is not numerical. With EM, 
the output is drawn from a probability distribution chosen based on a utility function. Consider the generalized problem where for an input $d$, output $s$ is chosen from the space denoted by $S$, i.e. $s\in S$. The utility function $u$ takes as input two parameters $d$ and $s$, and returns a real value $r$ measuring the quality of $s$ as a solution for the input $x$. EM aims to determine in a differentially private way $max_{s\in S}\{u(x,s) \}$. In general a single record may have a significant impact on the utility, hence the required noise may grow large, leading to poor accuracy.

\subsection{Problem Formulation}\label{Sec: problem formulation}

Consider a two-dimensional location dataset $D$ discretized to an arbitrarily-fine $N\times M$ grid. Each point is represented by its corresponding {\em rectangular cell} in the grid. We study the problem of releasing DP-compliant {\em histograms} to answer count queries as accurately as possible. Cell counts are modeled via an $N\times M$ {\em frequency matrix}, in which the entry in the $i^{th}$ row and $j^{th}$ column represents the number of records located inside cell $(i,j)$ of the grid. 

A DP histogram is generated based on a non-overlapping partitioning of the frequency matrix by applying methods to preserve $\epsilon$-DP. The DP histogram consists of the boundary of {\em partitions} and their noisy counts, where each partition consists of a group of cells. 

Let us denote the total count of a partition with $q$ cells by $c$ and its noisy count by $\overline{c}$. There are two sources of error in answering a query. The first is referred to as {\em noise error}, which is due to Laplace noise added to the partition count. 
The second source of noise is referred to as {\em uniformity error}  and arises when a query has partial overlap with a partition. An assumption of uniformity is made within the partition, and the answer {\em per cell} is calculated as $\overline{c}/q$.

\begin{figure}[t]
	\subfloat[\label{Fig: Overview-a}]{%
	\includegraphics[scale=.45]{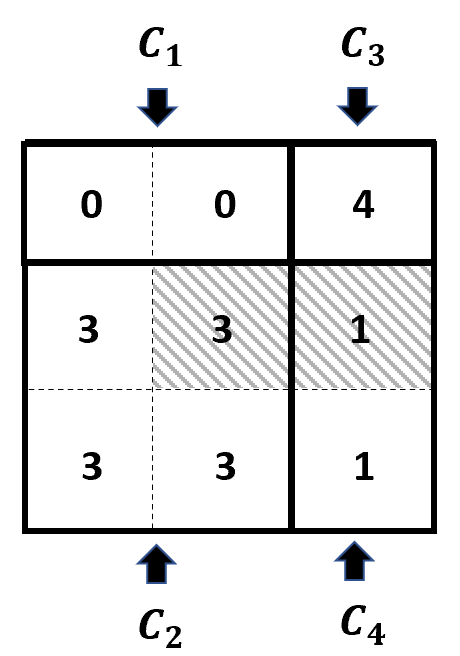}
	}
	\hfill
	\subfloat[\label{Fig: Overview-b}]{%
	\includegraphics[scale=.45]{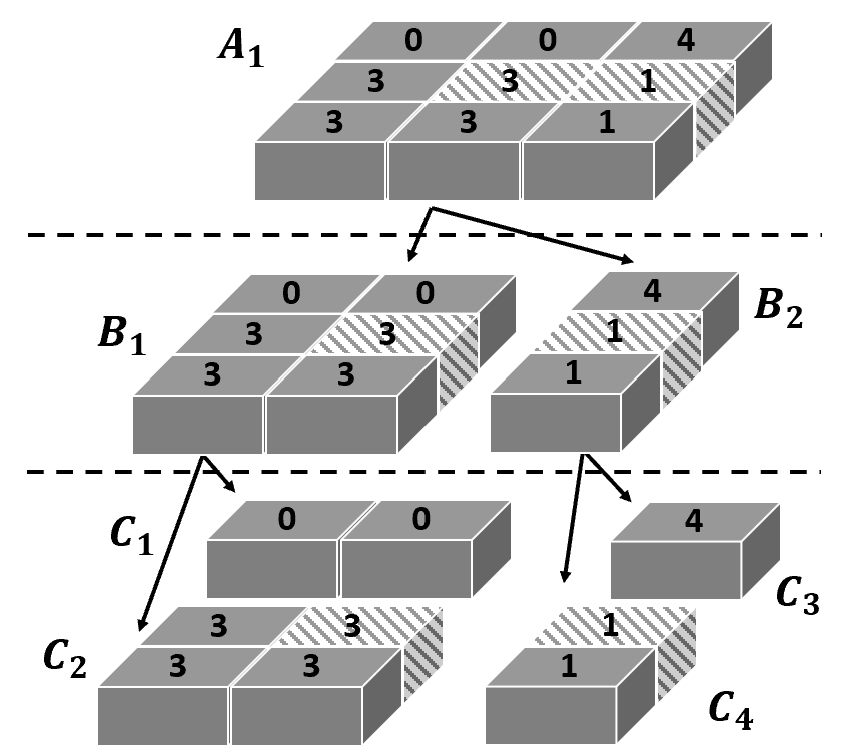}
	}
\centering
\vspace{-10pt}
\caption{Example of HTF partitioning. The dashed rectangles show the query.}
\label{Fig: Overview}
\vspace{-20pt}
\end{figure}

For example, consider the $3\times 3$ grid shown in Fig.~\ref{Fig: Overview-a}, where each count represents the number of data points in the corresponding cell. The cells are grouped in four partitions $C_1$, $C_2$, $C_3$, and $C_4$, each entailing $0,\, 12,\, 4$ and $2$ data points, respectively. Independent noise with the same magnitude is added to each partition's count denoted by $n_1$, $n_2$, $n_3$, and $n_4$, and released to the public as a DP histogram. The result of the query shown by the dashed rectangle can be calculated as $(12+n_2)/4 + (2+n_4)/2$.




\begin{problem}\label{problem statement}
Generate a DP histogram of dataset $D$, such that the expected value of relative error (MRE) is minimized, where for a query $q$ with the true count $c$ and noisy count $\overline{c}$ RE is calculated as 
\begin{equation}
   MRE(q) =  \dfrac{|c- \overline{c}|}{c}\times 100
\end{equation}
\end{problem}


In the past, several approaches have been developed for Problem~\ref{problem statement}. Still, current solutions have poor accuracy, which  limits their practicality. Some methods tend to perform better when applied to specific datasets (e.g., uniform) and quite poorly when applied to others. Limitations of existing work have been thoroughly evaluated in~\cite{ashwin}, and we review them in Section~\ref{Sec: literature reivew}.

\section{Homogeneous-Tree Framework}

Our proposed approach relies on two key observations to reduce the noise error and uniformity error. To address {\em noise error}, one needs to carefully calibrate the sensitivity of each operation performed, in order to reduce the magnitude of required noise. We achieve this objective by carefully controlling the depth of the indexing structure. To control the impact of {\em uniformity error}, we guide our structure-construction algorithm such that each resulting partition (i.e., internal node or leaf node) has a homogeneous data distribution within its boundaries. 

Homogeneity ensures that uniformity error is minimized, since a query that does not perfectly align with the boundaries of an internal/leaf node is answered by scaling the count within that node in proportion with the overlap between the query and the node. None of the existing works on DP-compliant data structures has directly addressed homogeneity. Furthermore, conventional spatial indexing structures (designed for non-private data access) are typically designed to optimize criteria other than homogeneity (e.g., reduce node area or perimeter, control the data count balance across nodes). As a result, existing approaches that use such structures  underperform when DP-compliant noise is present.

We propose a {\em Homogeneous-Tree Framework (HTF)} which builds a customized spatial index structure specifically designed for DP-compliant releases. We address directly aspects such as selection of structure height, a homogeneity-driven node split strategy, and careful tuning of privacy budget for each structure level. Our proposed data structure shares similarities with kd-trees, due to the specific requirements of DP: namely, (1) nodes should not overlap, since that would result in increased sensitivity, and (2) the leaf set should cover the entire data domain, such that an adversary cannot exclude specific areas by inspecting node boundaries. However, as shown in previous work, using kd-trees directly for DP releases leads to poor accuracy~\cite{cormode2012differentially,ashwin}.


Similar to kd-trees, HTF divides a node into two sub-regions across a split dimension, which is alternated through successive levels of the tree. The root node covers the whole dataspace. Figure~\ref{Fig: Overview-b} provides an example of a non-private simplified version of the proposed HTF construction applied on a $3\times 3$ grid (frequency matrix).
HTF consists of three steps:

(A) {\em Space partitioning} aims to find an enhanced partitioning of the map such that the accuracy of the private histogram is maximized. HTF performs heuristic partitioning based on a {\em homogeneity metric} we define. At every split, we choose the coordinate that results in the highest value of the homogeneity metric. For example, in the running example (Fig.~\ref{Fig: Overview-b}) node $B_1$ is split into $C_1$ and $C_2$, which are homogeneous partitions. However, the metric evaluation is not straightforward in the private case, as metric computation for each candidate split point consumes privacy budget. We use the Laplace mechanism to determine an advantageous split point without consuming large amounts of budget. As part of HTF, a search mechanism is used to select plausible candidates for evaluation and find a near-optimal split position. The total privacy budget allocated for the private partitioning is denoted by $\epsilon_{\text{prt}}$.      

(B)  {\em Data sanitization} starts by traversing the tree generated in the partitioning step. At each node, a certain amount of budget is used to perturb the node count using the Laplace mechanism. Based on the sanitized count, HTF evaluates the {\em stop condition} (i.e., whether to follow the downstream path from that node or release it as is), which is an important aspect in building private data structures. The private evaluation of stop conditions enables HTF to avoid over- or under-partitioning of the space, and preserve good accuracy. Revisiting the example in Fig.~\ref{Fig: Overview-b}, suppose that we do not want to further partition the space when the number of data points in a node is less than $7$. Once HTF reaches node $B_2$, the actual node count ($6$) is noise-perturbed. The value of the sanitized count may be less than $7$ after sanitization, leading to pruning at $B_2$ and stopping further partitioning. 
Finally, the tree's leaf set (i.e., sanitized count of each leaf node) is released to the public. The total budget used for data sanitization is denoted by $\epsilon_{\text{data}}$. 

(C) {\em Height estimation} is another important HTF step. Tree height is an important factor in improving accuracy, as it influences the budget allocated at each index level. HTF dedicates a relatively small amount of budget ($\epsilon_{\text{height}}$) to determine an appropriate height.

The total budget consumption of HTF ($\epsilon_{\text{tot}}$) is the sum of budgets used in each of the three steps: 
\begin{equation}
    \epsilon_{\text{tot}} = \epsilon_{\text{prt}} + \epsilon_{\text{data}}+ \epsilon_{\text{height}}
\end{equation}
The DP composition rules in the case of HTF apply as follows:
\begin{itemize}
    \item {\em Sequential decomposition:} The sum of budgets used for node splits along every tree path adds up to the total budget available for partitioning.
    \item {\em Parallel decomposition:} The budget allocated for partitioning nodes in the same level is independent, since the nodes at the same level have non-overlapping extents.
\end{itemize}

\section{Technical Approach}

Section~\ref{Sec: Objective Function} introduces the split objective function used in HTF, and provides its sensitivity analysis.
Section~\ref{Sec: DP Partitioning} focuses on HTF index structure construction.   Section~\ref{Sec: Data Perturbation} presents the data perturbation algorithm used to protect leaf node counts.

\subsection{Homogeneity-based Partitioning}\label{Sec: Objective Function}

Previous approaches that used kd-tree variations for DP-compliant indexes preserved the original split heuristics of the kd-tree: namely, node splits were performed on either median or average values of the enclosed data points. To preserve DP, the split positions were computed using the exponential mechanism (Section 2) which computes a merit function for each candidate split. 
However, such an approach results in poor query accuracy~\cite{ashwin}.


We propose {\em homogeneity} as the key factor for guiding splits in the HTF index structure. This decision is based on the observation that if all data points are uniformly distributed within a node, then the uniformity error that results when intersecting that node with the query range is minimized. 
At each index node split, we aim to obtain two new nodes with a high degree of intra-node density homogeneity. Of course, since the decision is data-dependent, the split point must be computed in a DP-compliant fashion. 

For a given node of the tree, suppose that the corresponding partition covers $U\times V$ cells of the $N\times M$ grid (i.e., frequency matrix), in which the count of data points located in its $i^{th}$ row and $j^{th}$ column is denoted by $c_{ij}$. Without loss of generality, we discuss the partitioning method w.r.t. the horizontal axis (i.e., rows). The aim is to find an index $k$ which groups rows $1$ to $k$ into one node and rows $k+1$ to $U$ into another, such that homogeneity is maximized within each of the {\em resulting} nodes (we also refer to resulting nodes as {\em clusters}).  We emphasize that the input grid abstraction is used in order to obtain a finite set of candidate split points. This is different than alternate approaches that use grids to obtain DP-compliant releases. Furthermore, the frequency matrix can be arbitrarily fined-grained, so discretization does not impose a significant constraint.  

The proposed split objective function is formally defined as:

\begin{align}\label{Equ: objective function}
    o_{k} = \sum_{i\Equal1}^k \sum_{j\Equal1}^V |c_{ij} - \mu_1| + \sum_{i\Equal k+1}^U \sum_{j\Equal1}^V |c_{ij}  - \mu_2|,
\end{align}
where
\begin{equation}
    \mu_1 = \dfrac{\sum_{i\Equal1}^k \sum_{j\Equal1}^V c_{ij} }{k\times V},\;\;\;\;\;\;  
    \mu_2 = \dfrac{\sum_{i\Equal k+1}^U \sum_{j\Equal1}^V c_{ij}}{(U-k)\times V}.
\end{equation}

The optimal index $k^*$  minimizes the objective function.
\begin{equation}
   k^* = \arg\min_{k}\;\; o_{k} \;\;\; 
\end{equation}

Consider the example in Figure~\ref{Fig: Overview-b} and the partitioning conducted for node $B_1$. There exist three possible ways to split rows of the frequency matrix: (i) separate the top row of cells resulting in clusters \{[0,0]\} and \{[3,3],[3,3]\} yielding the objective value of zero in Eq.~\eqref{Equ: objective function}; (ii) separate the bottom row of cells resulting in two clusters  \{[0,0],[3,3]\} and \{[3,3]\} yielding the objective value of $6$, or (iii) no  division is performed, yielding the objective value of $8$. Therefore, the proposed algorithm will select the first option ($k^*\Equal 2$), generating two nodes $C_1$ and $C_2$.

Note that the value of $k^*$ is not private, since individual location data were used in the process of calculating the optimal index. Hence, a DP mechanism is required to preserve privacy. Thus, we need to assess the sensitivity of $k^*$, which represents the maximum change in the split coordinate that can occur when adding or removing a single data point. The sensitivity calculation is not trivial, since a single data point can cause the optimal split to shift to a new position far from the non-private value. Another challenge is that the exponential mechanism, commonly used in literature to select candidates from a set based on a cost function, tends to have high sensitivity, resulting in low accuracy. 

\subsubsection{Baseline Split Point Selection.}

We propose a DP-compliant homogeneity-driven split point selection technique based on the Laplace mechanism. As before, consider $U\times V$ frequency matrix of a given node and a horizontal dimension split. Denote by $o_k$ the objective function for split coordinate $k$ among the $U$ candidates. There are $U$ possible outputs $\mathcal{O} = (o_{1},o_{2},...,o_{U})$, one for each split candidate. In a non-private setting, the index corresponding to the minimum $o_{i}$ value would be chosen as the optimal division. To preserve DP, given that the partitioning budget per computation is $\epsilon_{\text{prt}}''$, we add independent Laplace noise to each $o_{i}$, and then select the optimal value among all noisy outputs.
\begin{equation}
   \overline{\mathcal{O}} = (\overline{o}_{1},\overline{o}_{2},...,\overline{o}_{U}) = \mathcal{O} + \textbf{Lap}(2/\epsilon_{\text{prt}}''), 
\end{equation}
where $\textbf{Lap}(2/\epsilon_{\text{prt}}'')$ denotes a tuple of $U$ independent samples of Laplace noise. Note that since the grid is fixed, enumerating split candidates as cell coordinates is data-independent, hence does not incur disclosure risk. 
The Laplace noise added to each $o_i$ is calibrated according to a sensitivity of 2, as proved in Theorem~\ref{thm: sensitivity} (proof in Appendix A):
\begin{thm}[Sensitivity of Partitioning]\label{thm: sensitivity}
    The sensitivity of cost function $o_k$ for any given horizontal or vertical index $k$ is $2$.
\end{thm}
We refer to the above approach as the {\em baseline} approach. One challenge with the baseline is that the calculation of noise is performed separately for each candidate split point, and since the computation depends on all data points within the parent node, the budget consumption adds up according to {\em sequential composition}. 
This means that the calculation of each individual split candidate in $\overline{o_i}$ may receive only $1/U$ of the budget available for that level.

For large values of $U$, the privacy budget per computation becomes too small, decreasing accuracy. This leads to an interesting trade-off between the number of split point candidates evaluated and the accuracy of the entire release. On one hand, increasing the number of candidates leads to a higher likelihood of including the optimal split coordinate in the set $\overline{\mathcal{O}}$; on the other hand, there will be more noise added to each candidate's objective function output, leading to the selection of a sub-optimal candidate. Next, we propose an optimization which finds a good compromise between number of candidates and privacy budget per candidate. 

\subsubsection{Optimized Split Point Selection}

We propose an optimization that aims to minimize the number of split point candidate evaluations required, and searches for a local minimum rather than the global one. Algorithm~\ref{Algo: estimator} outlines the approach for a single split step along the $y$-axis (i.e., row split). 
Inputs to Algorithm~\ref{Algo: estimator} include (i) the frequency matrix $F_{U\times V}$ of the parent node, (ii) the total budget allocated for the partitioning per level of the tree $\epsilon_{\text{prt}}'$, and (iii) variable $T$ which bounds the maximum number of objective function computations -- a key factor indicating the extent of search, and thus of the budget per operation. The proposed approach is essentially a search tree, determining the candidate split to minimize the objective function's output. The search starts from a wide range of candidates and narrows down within each interval until reaching a local minimum, similar to a binary search.

Let $\{l,\ldots,r\}$ represent the index range where the search is conducted, initially set to the first and last possible index of the input frequency matrix. At every iteration of the main 'for' loop, the search interval is divided into four equal length sub-intervals, including three {\em inner} points and two {\em boundary} point. The inner points are referred to as {\em split indices}. The objective function is calculated for each of these candidates, and perturbed using Laplace noise to satisfy DP. The split corresponding to the minimum value is chosen as the center of the next search interval, and its immediate 'before' and 'after' split positions are assigned as the updated search boundaries $l$ and $r$). Hence, in every iteration, two new computations of the objective function are performed, except the first run which has a single computation. Therefore, the total number of private evaluations sums to $(2T+1)$ each perturbed with the privacy budget of $\epsilon_{\text{prt}}'' = \epsilon_{\text{prt}}'/(2T+1)$. 


\begin{algorithm}[t]
\caption{Near-optimal Split Point Estimator}\label{Algo: estimator}
\begin{algorithmic}[1]
\Function{GetSplitPoint}{ $F_{U\times V}$, $\epsilon_\text{prt}'$, $axis$, $T$ } 
\State   $\epsilon_{\text{prt}}'' \leftarrow \epsilon_{\text{prt}}'/(2\, T+1)$ 
\State $l\leftarrow 1,\, r\leftarrow\, (axis==0) ? V : U$ \#$axis=0$ means $x$-split
\State $k\leftarrow \lfloor (r - l)/2\rfloor$ 
\State Compute $o_{k}$ at $axis$ according to Eq. (\ref{Equ: objective function})
\State  $\overline{o}_{k}$ $\leftarrow \,o_{k}$ +   $\text{Lap}(2/\epsilon_{\text{prt}}'')$ 
\While {$l \leq r$ and $T > 0$}
    \State $k_1 \leftarrow \lfloor (k - l)/2\rfloor$
    \State $k_2\leftarrow \lfloor (r - k)/2\rfloor$
\label{Comp_start}
    \State Compute $o_{k1}$ and $o_{k2}$ at $axis$ according to Eq. (\ref{Equ: objective function})
    \State ( $\overline{o}_{k1}$, $\overline{o}_{k2}$ ) $\leftarrow (\,o_{k1}$, $o_{k2}\,$) +   $\textbf{Lap}(2/\epsilon_{\text{prt}}'')$  
\State $MinOutput\leftarrow min(\overline{{o_k}_{1}},\,  \overline{o_k},\, \overline{{o_k}_{2}})$
\If{$MinOutput = \overline{o_k}$}
    \State $l \leftarrow k_1,\, r\leftarrow k_2$
\ElsIf{$MinOutput = \overline{{o_k}_{1}}$}
    \State $k \leftarrow k_1,\, r\leftarrow  k $
\Else 
    \State $l \leftarrow k,\, k\leftarrow k_2$
\EndIf 
\State $T \leftarrow T - 1$
\EndWhile
\State \textbf{return} $k$
\EndFunction
\end{algorithmic}
\end{algorithm}

\subsection{HTF Index Structure Construction}\label{Sec: DP Partitioning}

Our proposed HTF index structure is built in accordance to the split point selection algorithm introduced in Section~\ref{Sec: Objective Function}.
The HTF construction pseudocode is presented in Algorithm~\ref{Algo: paritioning}. 
Each node stores the rectangular spatial extent of the node ($node$.region), its children ($node$.left and  $node$.right), real data count ($node$.count), noisy count ($node$.ncount), and the node's height in the tree.

The root of the tree represents the entire data domain ($N\times M$ frequency matrix) and its height is denoted by $h$. Deciding the height of the tree is a challenging task: a large height will result in a smaller amount of privacy budget per level, whereas a small one does not provide sufficient granularity at the leaf level, decreasing query precision.
We estimate an advantageous height value using a small amount of budget ($\epsilon_{\text{height}}$) to perturb the total number of data records based on the Laplace mechanism:

\begin{equation}
    \overline{|D|} = |D| + Lap(1/\epsilon_{\text{height}}).
\end{equation}
Next, we set the height to:

\begin{equation}
     h = \log_2{(\dfrac{\overline{|D|} \epsilon_{\text{tot}}}{10})}
\end{equation}

The formula is motivated by the work in~\cite{AG}. The authors show that when data are uniformly distributed in space, using a grid with a lower granularity of $\sqrt{\dfrac{\overline{|D|} \epsilon_{\text{tot}}}{c_0}}\times\sqrt{\dfrac{\overline{|D|} \epsilon_{\text{tot}}}{c_0}}$ improves the mean relative error, where the value of constant $c_0$ is set to 10 experimentally. We emphasize that the approach does not indicate that the number of leaves on the tree is $\dfrac{\overline{|D|} \epsilon_{\text{tot}}}{c_0}$, but the information contained in this number is merely used as an estimator of the tree's height. 
This estimation is formally characterized in~\cite{ashwin} and referred to as {\em scale-epsilon exchangeability property}. The intuition is that the error due to decreasing the amount of budget used for the estimation is offset by having a larger number of data points in the entire dataset.

The last input to the algorithm is the budget allocated per level of the partitioning tree. We use {\em uniform budget allocation} to allocate the budget between levels denoted as $\epsilon_{\text{prt}}'= \epsilon_{\text{prt}}/h$.

Starting from the root node, the proposed algorithm recursively creates two child nodes and decreases height by one. This is done by splitting the underlying area of the node into two hyperplanes based on Algorithm~\ref{Algo: estimator}. The division is done on the $y$ dimension if the current height is an even number and in the $x$ dimension otherwise. The algorithm continues until reaching the minimum height of zero, or to a point where no further splitting is possible. 

\begin{algorithm}[t]
\caption{DP space partitioning}\label{Algo: paritioning}
\begin{algorithmic}[1]
\Function{PrivatePartitioning}{{\em node}, $\epsilon_{\text{prt}}'$} 
    \If {$node$.height=0}
    \State \textbf{return} $node$
    \EndIf
\State $axis\leftarrow$ $node$.height \textbf{mod} 2 
\State $node.count\leftarrow sum(node.\text{region})$
\State $OptIdx\leftarrow $ GetSplitPoint($node$.FreqMatrix, $\epsilon_{\text{prt}}'$, axis, T ) 
\State $leftC$, $rightC$ $\leftarrow$ split $node$ on $OptIdx$
\State $node$.leftChild $\leftarrow$ PrivatePartitioning($leftChild$,  $\epsilon_{\text{prt}}'$)   
\State $node$.rightChild $\leftarrow$ PrivatePartitioning($rightChild$, $\epsilon_{\text{prt}}'$)
\EndFunction
\end{algorithmic}
\end{algorithm}

\subsection{Leaf Node Count Perturbation}\label{Sec: Data Perturbation}

Once the HTF structure is completed, the final step of our algorithm is to release DP-compliant counts for index nodes, so that answers to queries can be reconstructed from the noisy counts. The total partitioning budget adds up to $\epsilon_{\text{height}}+ \epsilon_{\text{prt}}$, where $\epsilon_{\text{height}}$ was used to estimate the tree height and $\epsilon_{\text{prt}}$ budget to generate the private partitioning tree. The data perturbation step uses the remaining $\epsilon_{\text{data}}$ amount of budget and releases node counts according to the Laplace mechanism.


One can choose various strategies to release index node counts. At one extreme, one can simply release a noisy count for each index node; in this case, the budget must be shared across nodes on the same path (sequential composition), and can be re-used across different paths (parallel composition). This approach has the advantage of simplicity, and may do well when queries exhibit large variance in size -- it is well-understood that when perturbing large counts, the relative error is much lower, since the Laplace noise magnitude only depends on sensitivity, and not the actual count.

However, in practice, queries tend to be more localized, and one may want to allocate more budget to the lower levels of the structure, where the actual counts are smaller, thus decreasing relative error. In fact, as another extreme, one can concentrate the entire $\epsilon_{\text{data}}$ on the leaf level. However, doing so can also decrease accuracy, since some leaf nodes have very small real counts.

Our approach takes a middle ground, where the available $\epsilon_{\text{data}}$ is spent to (i) determine which nodes to publish and (ii) ensure sufficient budget remains for the noisy counts. Specifically, we publish only leaf nodes, but these are not the same leaves returned by the structure construction algorithm. Instead, we perform an additional pruning step which uses  the noisy counts of internal nodes to determine a {\em stop condition}, i.e., the level at which a node count is likely to be small enough that a further recursion along that path is not helpful to obtain good accuracy. Effectively, we perform pruning of the tree using a small fraction of the data budget, and then split the remaining budget among the non-pruned nodes along a path. This helps decrease the {\em effective height} of the tree across each path, and hence the resulting budget per level increases.

Next, we present in detail our approach that contributes two main ideas: (i) how to determine smart stop (or pruning) conditions based on noisy internal node counts, and (ii) how to allocate perturbation budget across shortened paths.


The proposed technique is summarized in Algorithm~\ref{Algo: data perturbation}: it takes as inputs the root node of the tree generated in the data partitioning step; the remaining budget allocated for the perturbation of data ($\epsilon_{\text{data}}$); a tracker of accumulated budget ($\epsilon_{\text{accu}}$); a stop condition predicate denoted by $cond$; and the nominal tree height $h$ as computed in Section~\ref{Sec: DP Partitioning}. Similar to prior work~\cite{cormode2012differentially}, we use a {\em geometric progression} budget allocation strategy, but we enhance it to avoid wasting budget on unnecessarily long paths. The intuition behind this strategy is to assign more budget to the nodes located in the lower levels of the tree, since their actual counts are lower, and hence larger added noise impacts the relative error disproportionately high. Conversely, at the higher levels of the tree, where actual counts are much higher, the effect of the noise is negligible.

Eq.~\eqref{Equ: geometric progression} formulates this goal as a convex optimization problem. 
\begin{align}\label{Equ: geometric progression}
    \underset{\epsilon_0...\epsilon_h}{min}\;\;\;& \sum_{i\Equal0}^h 2^{h-i}/\epsilon_i^2\\
    \text{where} \\
    & \sum_{i\Equal0}^h \epsilon_i=\epsilon, \;\; \epsilon_i>0 \; \;\forall i=0...h
\end{align}
Writing Karush-Kuhn-Tucker (KKT)~\cite{boyd2004convex} conditions, the optimal allocation of budget can be calculated as:

\begin{align}
    L(\epsilon_1,...,\epsilon_h,\lambda) &= \sum_{i\Equal0}^h 2^{h-i}/\epsilon_i^2 + \lambda ( \sum_{i\Equal0}^h \epsilon_i- \epsilon)\\
    &\Rightarrow \dfrac{\partial L}{\partial\epsilon_i} = - \dfrac{2^{h-i+1}}{\epsilon_i^3}+ \lambda =0\\
    &\Rightarrow \epsilon_i = \dfrac{2^{h-i+1}}{\lambda^{1/3}},
\end{align}
and substituting $\epsilon_i$'s in the constraint of problem the optimal budget in the $i$-th level is derived as
\begin{equation}
    \epsilon_i = \dfrac{2^{(h-i)/3}\,\epsilon\, (2^{1/3}-1)}{(2^{(h+1)/3}-1)}.
\end{equation}

The algorithm starts the traversal from the partitioning tree's root and recursively visits the descendent nodes. Once a new node is visited, the first step is to use the node's height to determine the allocated budget ($\epsilon_{\text{data}}'$) based on geometric progression. Recall that the nodes on the same level follow parallel decomposition of the budget as their underlying areas in space do not overlap. Additionally, the algorithm keeps track of the amount of budget used so far on the tree, optimizing the budget in later stages. 
Next, the computed value of $\epsilon_{\text{data}}'$ is utilized to perturb the $node$.count by adding Laplace noise, resulting in noisy count $node$.ncount. 

\begin{algorithm}[t]
\caption{DP data perturbation}\label{Algo: data perturbation}
\begin{algorithmic}[1]
\Function{Perturber}{{\em node}, $\epsilon_{\text{data}}$, $\epsilon_{\text{accu}}$ ,$cond$, $h$}
    \State $i\leftarrow node.\text{height}$
    \State $\epsilon_{\text{data}}'\leftarrow \dfrac{2^{(h-i)/3}\,\epsilon_{\text{data}}\, (2^{1/3}-1)}{(2^{(h+1)/3}-1)}$  
    \State $\epsilon_{\text{accu}} = \epsilon_{\text{accu}} + \epsilon_{\text{data}}'$
    \State $node$.ncount = $node$.count + $Lap(1/\epsilon_{\text{data}}')$
    \If{$node$.ncount$\leq cond$}
        
        \State $\epsilon_{\text{remain}} = \epsilon_{\text{data}} - \epsilon_{\text{accu}} $
        \State $node$.ncount = $node$.count + $Lap(1/\epsilon_{\text{remain}})$
        \State $node$.leftChild = $node$.rightChild = \textbf{null} 
    \Else    
        \State Perturber($node$.leftChild,\,$\epsilon_{\text{data}}$,\, $\epsilon_{\text{accu}}$,\,$cond$,\,$h$)
        \State Perturber($node$.rightChild ,\, $\epsilon_{\text{data}}$,\, $\epsilon_{\text{accu}}$,\,$cond$,$h$)
    \EndIf    
\EndFunction
\end{algorithmic}
\end{algorithm}

The stop condition we use takes into account the noisy count in the current internal node (i.e., count threshold); and the spatial extent of the internal node threshold (i.e., extent threshold). If none of the thresholds is met for the current node, the algorithm recursively visits the node's children; otherwise, the algorithm prunes the tree considering that the current node should be a leaf node. In the latter case, the algorithm subtracts the accumulated budget used so far on that path from the root, and uses the entire remaining budget available to perturb the count. This significantly improves the utility, as geometric allocation tends to save most of the budget for the lower levels of the tree. Revisiting the example in Figure~\ref{Fig: Overview-b}, suppose that the stop condition is to prune when the underlying area consists of less than four cells.  During the data perturbation process, the node $B_2$ is turned into a leaf node due to its low number of cells. At this point, the node's children are removed, and its noisy count is determined based on the remaining budget available on the lower levels of the tree.

\section{Experimental Evaluation}\label{Experimental Evaluation}

\subsection{Experimental Setup}

We evaluate HTF on both real and synthetic datasets: 

\noindent
    {\bf Los Angeles Dataset.} 
    This is a subset of the Veraset dataset~\cite{datarade}\footnote{Veraset is a data-as-a-service company that provides anonymized population movement data collected through location measurement signals of cell phones across USA.}, including location measurements of cell phones within LA city. In particular, we consider a large geographical region covering a $70 \times
70$ km$^2$ area centered at latitude 34.05223 and longitude -118.24368. The selected data generates a frequency matrix of 3.5 million data points during a time period between Jan 1-7 2020.  

\noindent
    {\bf Synthetic dataset.} We generate locations according to a Gaussian distribution as follows: a cluster center, denoted by $(x_c,y_c)$, is selected uniformly at random. Next, coordinates for each data point $x$ and $y$ are drawn from a Gaussian distribution with the mean of $x_c$ and $y_c$, respectively.  We model three sparsity levels by using three standard deviation ($\sigma$) settings for Gaussian variables: low ($\sigma = 20$), medium ($\sigma = 50$), and high ($\sigma = 100$) sparsity. 

We discretize the space to a $1024\times1024$ frequency matrix. 
We use as performance metric the mean relative error (MRE) for range queries. 
Similar to prior work~\cite{ashwin, zhang2016privtree, xiao2010differential, AG}, we consider a \textit{smoothing factor} of $20$ for the relative error, to deal with cases when the true count for a query is zero (i.e., relative error is not defined).
Each experimental run consists of 2,000 random rectangular queries with center selected uniformly at random. We vary the size of queries to a region covering $\{2\%, 6\%, 10\%\}$ of the dataspace.



\begin{figure*}[t]
	\subfloat[$\epsilon_{\text{tot}} = 0.1$, random shape and size queries.\label{DD1}]{%
	\includegraphics[scale=.6]{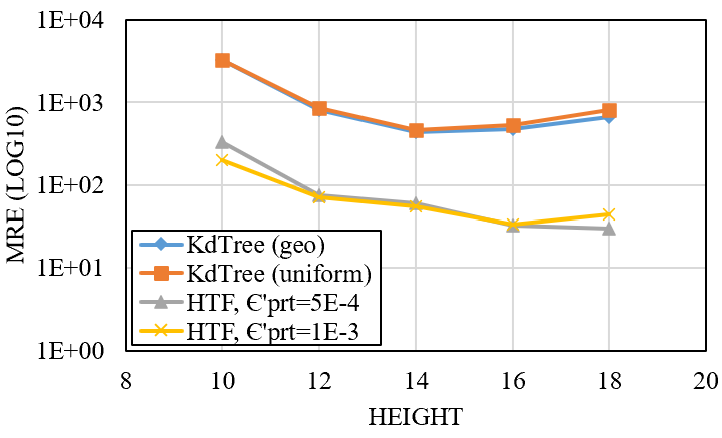}
	}
	\hfill
	\subfloat[$\epsilon_{\text{tot}} = 0.3$, random shape and size queries. \label{DD2}]{%
	\includegraphics[scale=.6]{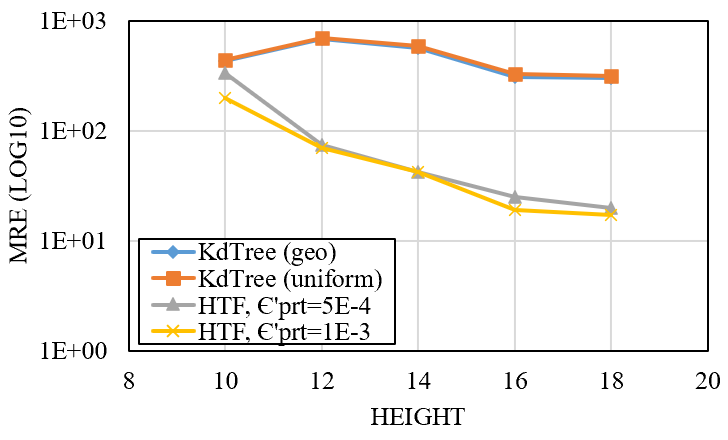}
	}
	\hfill
	\subfloat[$\epsilon_{\text{tot}} = 0.5$, random shape and size queries. \label{DD3}]{%
	\includegraphics[scale=.6]{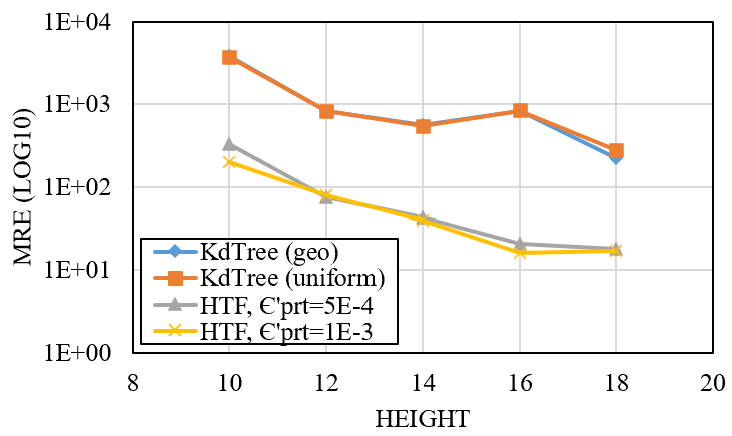}
	}
	\hfill
	\subfloat[Random square queries, size $2\%$, $\epsilon_{\text{tot}} = 0.1$.\label{DD4}]{%
	\includegraphics[scale=.53]{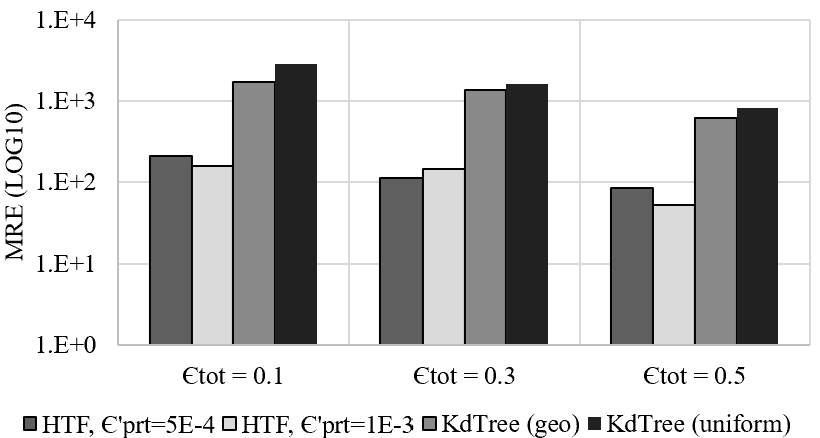}
	}
	\hfill
	\subfloat[Random square queries, size $6\%$, $\epsilon_{\text{tot}} = 0.1$.\label{DD5}]{%
	\includegraphics[scale=.53]{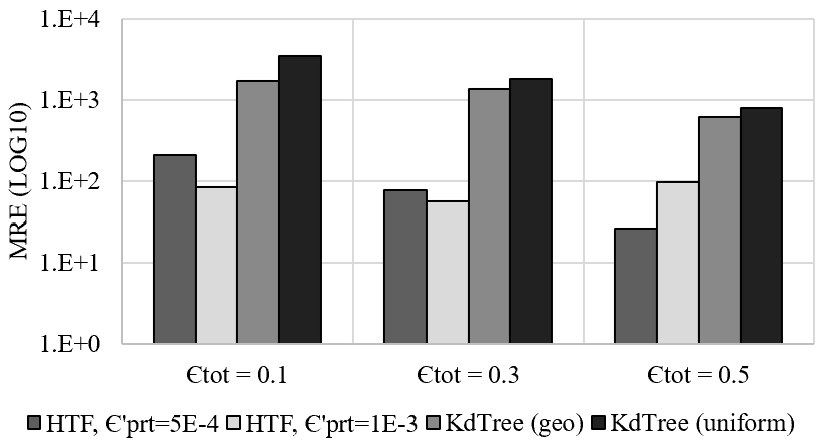}
	}
	\hfill
	\subfloat[Random square queries, size $10\%$, $\epsilon_{\text{tot}} = 0.1$.\label{DD6}]{%
	\includegraphics[scale=.53]{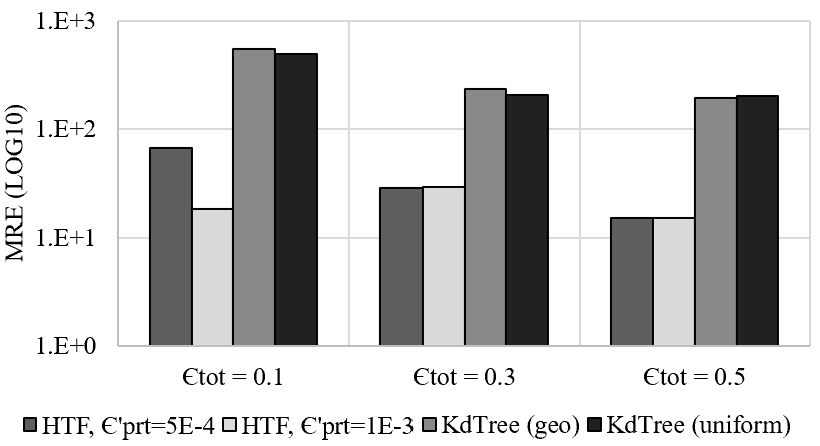}
	}
	\vspace{-10pt}
	\caption{Comparison with Data Dependent Algorithms, Los Angeles Dataset.}
	\label{Fig: DD comparison}
	\vspace{-10pt}
\end{figure*}

\subsection{HTF vs Data Dependent Algorithms}\label{Section: DD}

Data-dependent algorithms aim to exploit users' distribution to provide an enhanced partitioning of the map. 
The state-of-the-art data-dependent approach for DP-compliant location publication is the kd-tree technique from~\cite{cormode2012differentially}. The kd-tree algorithm generates the partitioning tree by splitting on median values, which are determined using the exponential mechanism. We have also included the smoothing post-processing step from \cite{AG, hay2009boosting} which resolves inconsistencies within the structure (e.g., using the fact that counts in a hierarchy should sum to the count of an ancestor).


Fig.~\ref{Fig: DD comparison} presents the comparison of the HTF algorithm with kd-tree approaches, namely: (i) geometric budget allocation in addition to smoothing and post-processing labelled as {\em KdTree (geo)}; (ii) uniform budget allocation including smoothing and post-processing labelled as {\em KdTree (uniform)}; (iii) HTF algorithm with the partitioning budget per level set to $\epsilon_{\text{prt}}' = 5E-4$; and (iv) HTF algorithm with $\epsilon_{\text{prt}}' = 1E-3$. Recall that $\epsilon_{\text{prt}}'$ denotes the budget per level of partitioning, and therefore, given the tree's height as $h$, the remaining budget for perturbation is derived as $\epsilon_{\text{data}} = \epsilon_{\text{tot}} - \epsilon_{\text{prt}}'\times h - \epsilon_{\text{height}} $. The value of $\epsilon_{\text{height}}$ in the experiments is set to $1E-4$, the HTF's $T$ value is set to $3$, and stop condition thresholds are set to no less than $5$ cells or $100$ data points. Moreover, for the kd-tree algorithm, $15\%$ of the total budget is allocated to implement the partitioning.

In Figs.~\ref{DD1},~\ref{DD2}, and \ref{DD3}, the MRE performance is compared for different values of $\epsilon_{\text{tot}}$ over a workload of uniformly located queries with random shape and size. HTF clearly outperforms kd-tree for all height settings. Looking at the MRE performance, the kd-tree algorithm follows a parabolic shape commonly occurring in tree-based algorithms, meaning that the MRE performance reaches its best values at a particular height, and further partitioning of the space increases the error. This is caused by excessive partitioning in low-density areas. HTF, on the other hand, is applying stop conditions to avoid the adverse effects of over-partitioning, and is able to estimate the optimal height beforehand. Figs.~\ref{DD4}, \ref{DD5}, and \ref{DD6}, show the results when varying query size (for square shape queries). HTF outperforms the kd-tree algorithm significantly. 

\begin{figure*}[t]
	\subfloat[Stop Count = 10, random shape/size queries.\label{PD1}]{%
	\includegraphics[scale=.55]{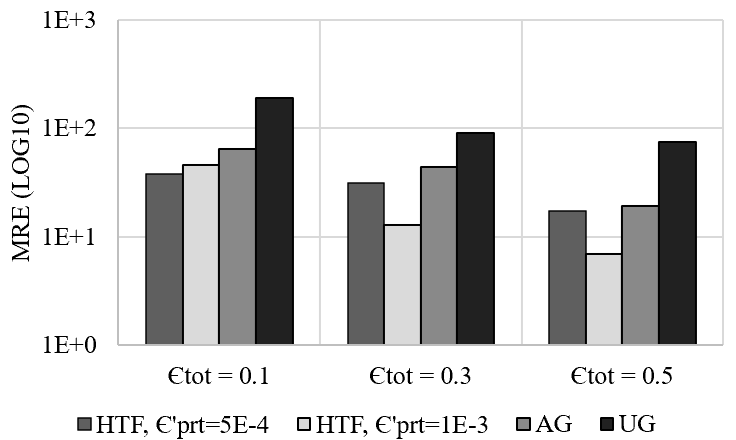}
	}
	\hfill
	\subfloat[Stop Count = 50, random shape/size queries. \label{PD2}]{%
	\includegraphics[scale=.55]{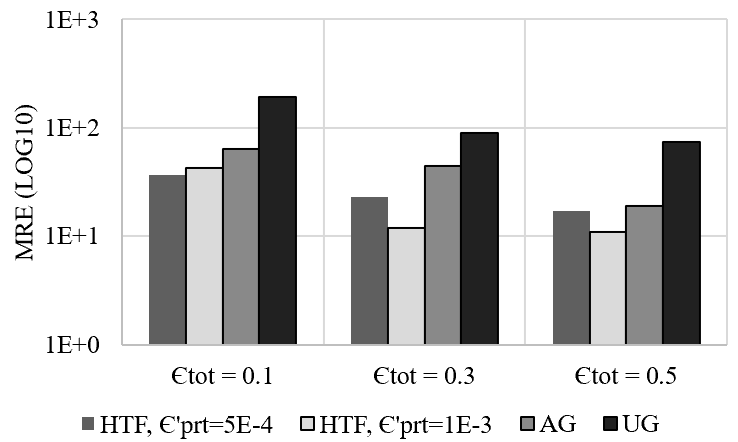}
	}
	\hfill
	\subfloat[Stop Count = 100, random shape/size queries \label{PD3}]{%
	\includegraphics[scale=.55]{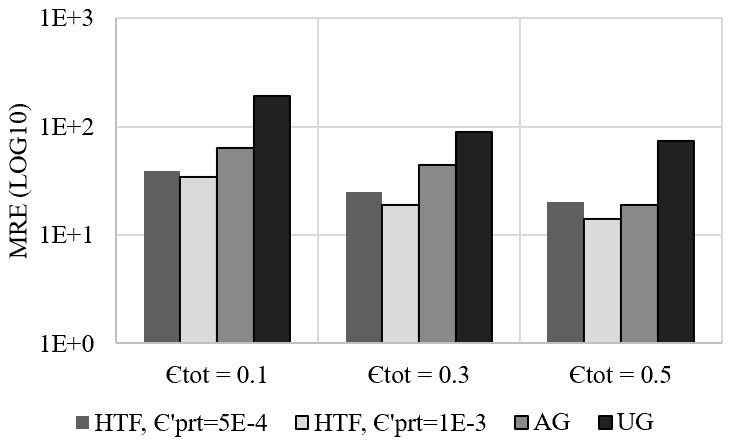}
	}
	\hfill
	\subfloat[Random square queries, size $2\%$.\label{PD4}]{%
	\includegraphics[scale=.55]{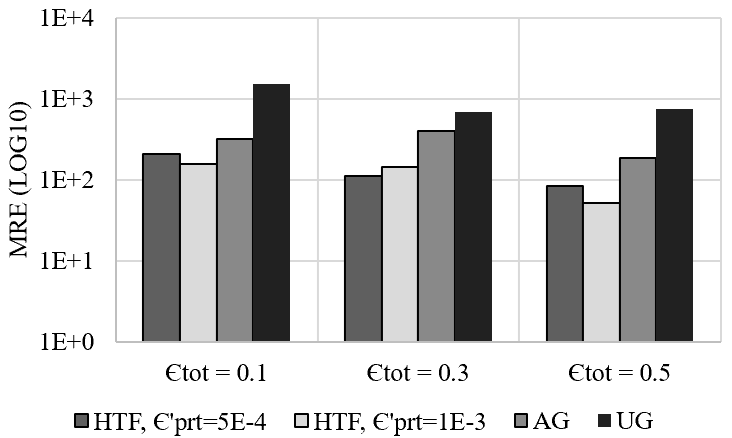}
	}
	\hfill
	\subfloat[Random square queries, size $6\%$.\label{PD5}]{%
	\includegraphics[scale=.55]{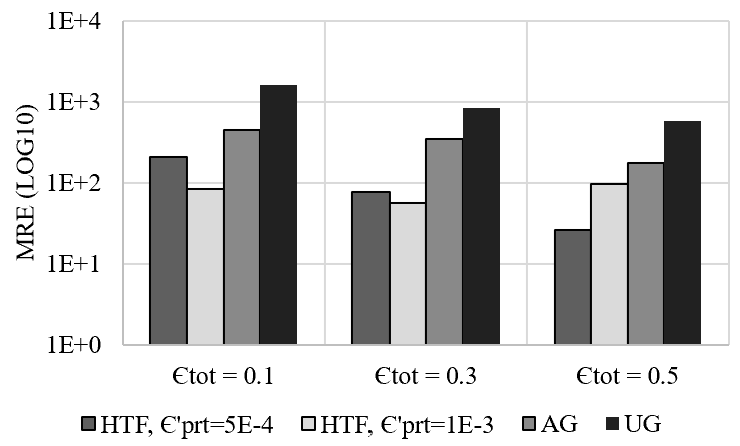}
	}
	\hfill
	\subfloat[Random square queries, size $10\%$.\label{PD6}]{%
	\includegraphics[scale=.55]{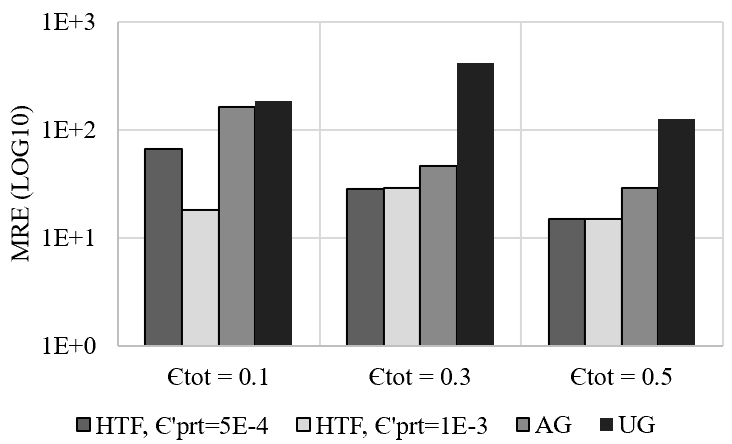}
	}
	\vspace{-10pt}
	\caption{Comparison to Grid-based Algorithms, Los Angeles Dataset.}
	\label{Fig: PD comparison}
	\vspace{-10pt}
\end{figure*}

\begin{figure*}[t]
	\subfloat[$\epsilon_{\text{tot}} = 0.1$, random shape and size queries.\label{DI1}]{%
	\includegraphics[scale=.55]{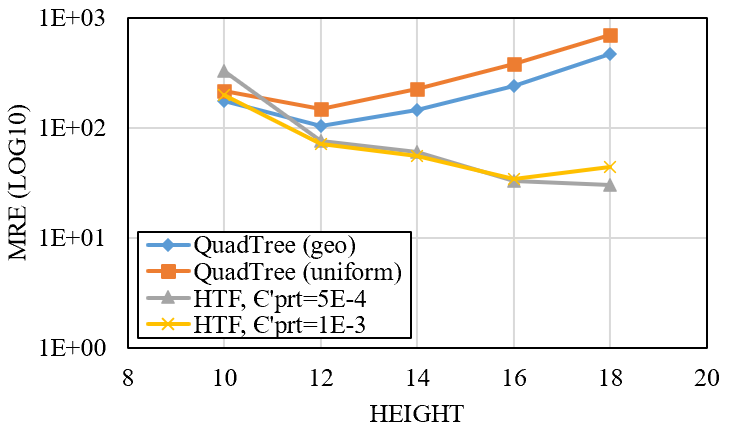}
	}
	\hfill
	\subfloat[$\epsilon_{\text{tot}} = 0.3$, random shape and size queries.\label{DI2}]{%
	\includegraphics[scale=.55]{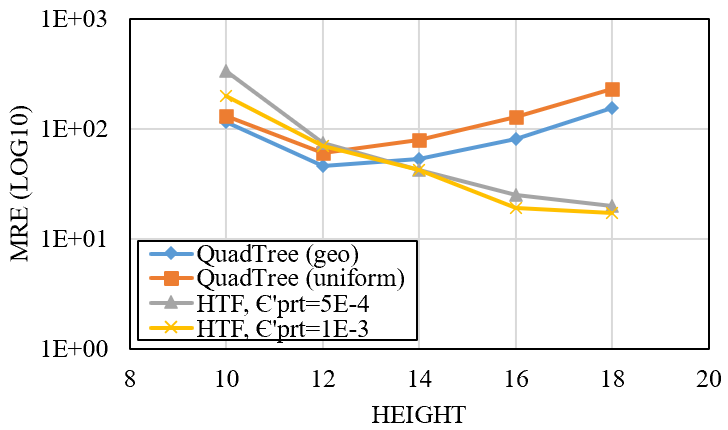}
	}
	\hfill
	\subfloat[$\epsilon_{\text{tot}} = 0.5$, random shape and size queries. \label{DI3}]{%
	\includegraphics[scale=.55]{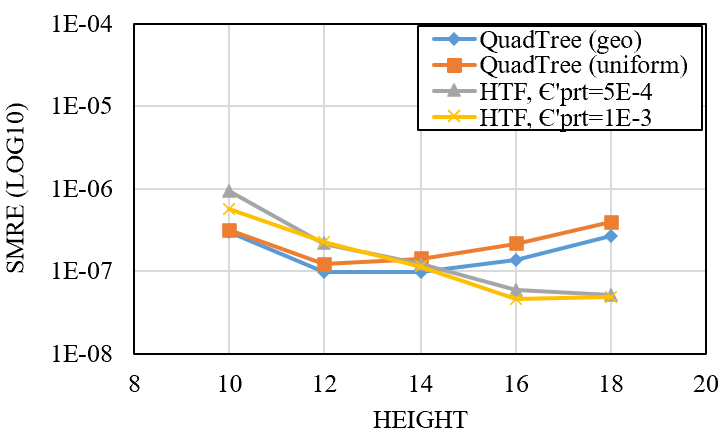}
	}
	\hfill
	\subfloat[Random square queries, size $2\%$. \label{DI4}]{%
	\includegraphics[scale=.49]{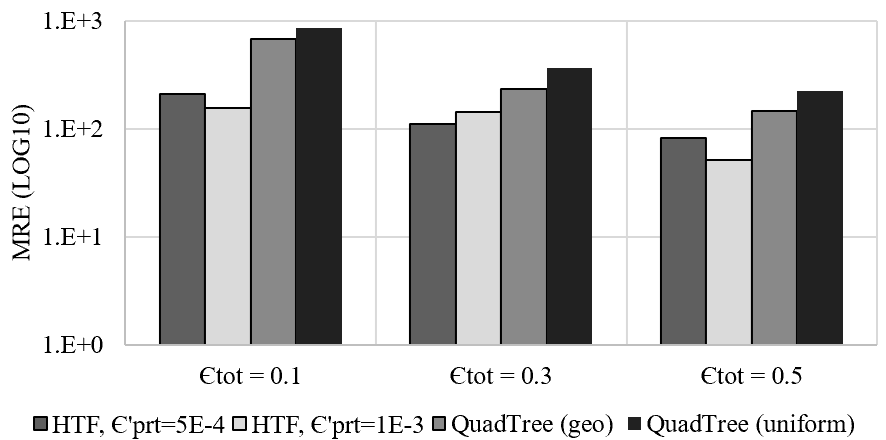}
	}
	\hfill
	\subfloat[Random square queries, size $6\%$. \label{DI5}]{%
	\includegraphics[scale=.49]{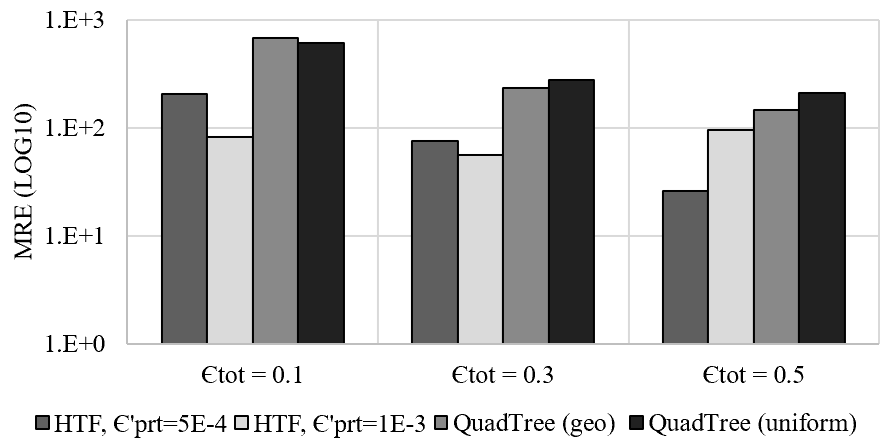}
	}
	\hfill
	\subfloat[Random square queries, size $10\%$.\label{DI6}]{%
	\includegraphics[scale=.49]{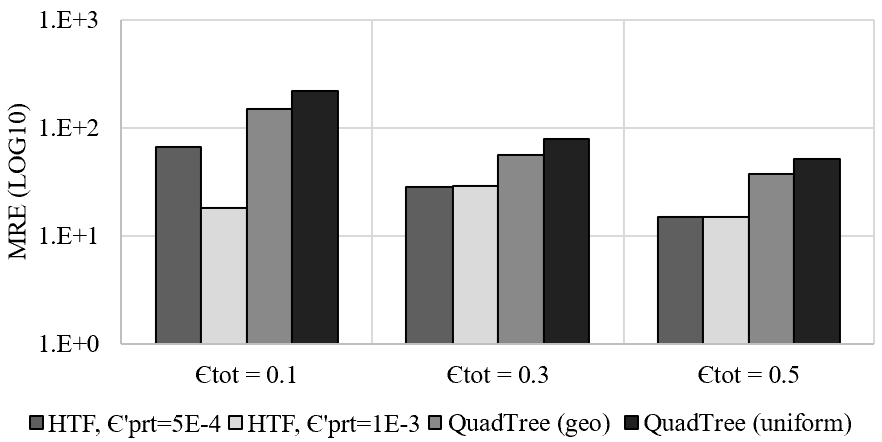}
	}
	\vspace{-10pt}
	\caption{Comparison to Data Independent Algorithms, Los Angeles Dataset.}
	\label{Fig: DI comparison}
	\vspace{-10pt}
\end{figure*}	

\subsection{HTF vs Grid-based Algorithms}\label{Section: PD}

Grid-based approaches are mostly data-independent, and they partition the space using regular grids. The uniform grid (UG) approach uses a single-layer fixed size grid, whereas its successor adaptive grid (AG) method considers two layers of regular grids: the first layer is similar to UG, whereas the second uses a small amount of data-dependent information (i.e., noisy query results on the first layer) to tune the second layer grid granularity.

Fig.~\ref{Fig: PD comparison} presents the comparison of HTF with AG and UG. For HTF, we consider several stop condition thresholds. HTF consistently outperforms grid-based approaches, especially when the total privacy budget is lower (i.e., more stringent privacy requirements). 
The impact of the stop count condition on HTF depends on the underlying distribution of data points. For the Los Angeles dataset, MRE is relatively larger for small values of $\epsilon_{\text{tot}}$. The performance improves and reaches its near-optimal values around stop count $50$, and ultimately worsens when the stop count becomes larger. This matches our expectation as, on one hand, small values of stop count result in over-partitioning, and on the other hand, when the stop count is too large, the partitioning tree cannot reach the ideal heights, resulting in high MRE values.

Figures~\ref{PD4}, \ref{PD5}, and \ref{PD6}, show the obtained results for varying query sizes ($2$, $6$, and $10\%$ of the data domain). HTF outperforms both AG and UG. Note that 
all three algorithms are adaptive and change their partitioning according to the number of data points. Therefore, in low privacy regimes, the structure of algorithms may cause fluctuations in the accuracy. However, as the privacy budget grows, the algorithms reach their maximum partitioning limit, and increasing the budget always results in lower MRE.

\subsection{HTF vs Data Independent Algorithms}\label{Section: DI}

The most prominent DP-compliant data-independent technique~\cite{cormode2012differentially} uses QuadTrees. 
 The technique recursively partitions the space into four equal size quadrants. Two commonly used budget allocation techniques used in~\cite{cormode2012differentially} are geometric budget allocation (geo) and uniform budget allocation. For a fair comparison, we have also included the smoothing post-processing step from~\cite{AG,xiao2012dpcube}.

Fig.~\ref{Fig: DI comparison} presents the comparison results. Figures~\ref{DI1}, \ref{DI2}, \ref{DI3} are generated using random shape and size queries, and several different height settings. Note that the fanout of QuadTrees is double the fanout of HTF, so the height represented by $2k$ in the figures corresponds to the implementation of QuadTree with the height of $k$. The error of the QuadTree approach is large for small heights, then improves to its optimal value, and rises again significantly as height further increases, due to over-partitioning. Similar to kd-trees, no systematic method has been developed for QuadTree to determine optimal height, whereas the HTF height selection heuristic yields levels $15$, $16$, and $17$ for the allocated privacy budget of $0.1$, $0.3$, and $0.5$, respectively. HTF outperforms QuadTree for all settings of $\epsilon_{\text{tot}}$.
Figures~\ref{DI4}, \ref{DI5}, and \ref{DI6} show the accuracy of HTF and QuadTree for square-shaped randomly placed queries of varying size. HTF outperforms Quadtree in all cases.

\subsection{Additional Benchmarks}\label{Section: validation}

\begin{figure}[h!]
\centering
\includegraphics[scale=.75]{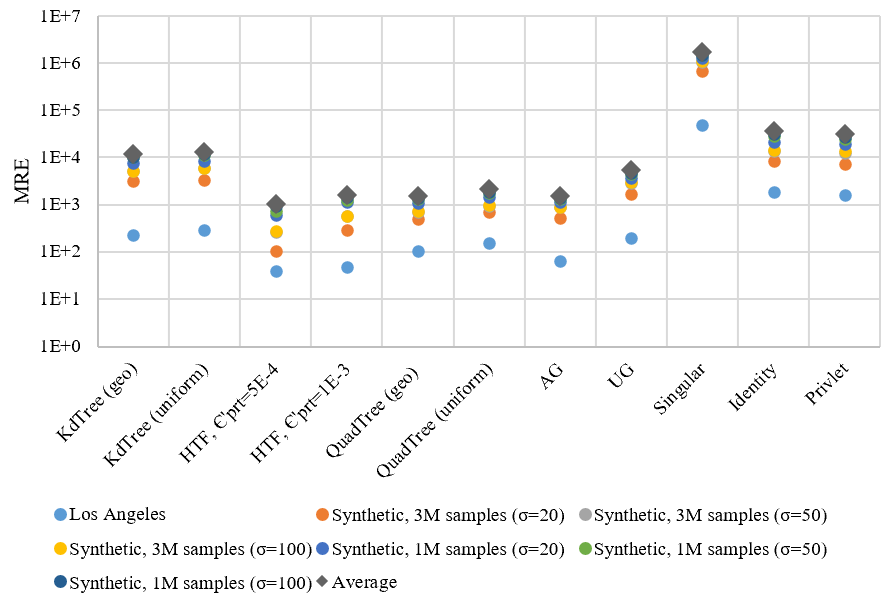}
\hspace{1em}
\centering
\vspace{-20pt}
\caption{Mixed workloads, $\epsilon_{\text{tot}}=0.1$, All Datasets.}
\label{Fig: All Algorithms}
\vspace{-15pt}
\end{figure}

To further validate HTF performance, we run experiments on the Los Angeles dataset as well as six synthetic datasets generated using Gaussian distribution. Fig.~\ref{Fig: All Algorithms} presents the comparison of all algorithms with randomly generated query workload and a privacy budget of $\epsilon_{\text{tot}}=0.1$. Three additional algorithms are used as comparison benchmarks, (i) {\em Singular} algorithm-- preserving differential privacy by adding independent Laplace noise to each entry of the frequency matrix, (ii) {\em Uniform} algorithm in which Laplace noise is added to the total count of the grid with the assumption that data are uniformly distributed within the grid, and (iii) the {\em Privlet}~\cite{xiao2010differential} algorithm based on wavelet transformations.

Fig.~\ref{Fig: All Algorithms} shows that HTF consistently outperforms existing approaches. For the denser datasets ($sigma=20$) the gain compared to approaches designed for uniform data (e.g., UG, AG, Quadtrees) is lower. As data sparsity grows, the difference in accuracy between HTF and the benchmarks increases. 
HTF performs best on a relative basis for lower  $\epsilon_{\text{prt}}'$, i.e., more stringent privacy requirements.


\vspace{-5pt}
\section{Related Work}\label{Sec: literature reivew}


The comprehensive benchmark analysis in~\cite{ashwin} showed that the dimensionality and scale of the data directly determine the accuracy of an algorithm. For one-dimensional data, both data-dependent and data-independent methods perform well. The hierarchical method in~\cite{hay2009boosting} uses a strategy consisting of hierarchically structured range queries arranged as a tree. Similar methods (e.g.,  \cite{zhang2016privtree}) differ in their approach to determining the tree's branching factor and allocating appropriate budget to each of its level. Data-dependent techniques on the other hand exploit correlation in real-world datasets in order to boost the accuracy of histograms. They first compress the data without loss: for example, EFPA \cite{acs2012differentially} applies the Discrete Fourier Transform,  whereas DAWA~\cite{li2014data} uses dynamic  programming to compute the least cost partitioning. The compressed data is then sanitized, for example, directly with LPM~\cite{acs2012differentially} or with a greedy algorithm that tunes the privacy budget to a sample query set given in advance~\cite{li2014data}. Privlet~ \cite{xiao2010differential} compresses data through a wavelet transformation such that the the noise incurred by a range query scales proportionately to the logarithm of its length.

In the 2D scenario, the main focus is on spatial datasets that exhibit sparse and skewed data distributions, where only data-dependent approaches tend to be competitive. General-purpose mechanisms such as the matrix mechanism of Li and Miklau \cite{li2013optimal, li2012adaptive} and its workload-aware counterpart DAWA \cite{li2014data} operate over a discrete 1D domain, and may be extended to spatial data by applying a Hilbert transform to the 2D data representation~\cite{ashwin}. However, approaches specialized for answering spatial range queries, such as UG~\cite{AG}, AG~\cite{AG}, QuadTree~\cite{cormode2012differentially} and kd-tree~\cite{xiao2010differentially} outperform general-purpose mechanisms~\cite{ashwin}. 
Xiao et al. \cite{xiao2010differentially} present the earliest attempt at a DP-compliant spatial decomposition algorithm based on the kd tree. It first imposes a uniform grid over the data domain, and then constructs a private kd tree over the cells in the grid. While the simplicity of the approach is appealing, the split criterion is solely based on the median, leading to high-sensitivity and split partitions with low intra-node homogeneity. 

Recent work focuses on high-dimensional data, where the key idea is to reduce the impact of the higher dimensionality. The most accurate algorithm in this class is High-Dimensional Matrix Mechanism (HDMM) \cite{mckenna2018optimizing} which represents queries and data as vectors and uses sophisticated optimization and inference techniques to answer them. DPCube~\cite{xiao2012dpcube} searches for dense sub-cubes to release privately. Some of the privacy budget is used to obtain noisy counts over a regular partitioning, which is then refined to a standard kd-tree. Fresh noisy counts for the partitions are obtained with the remaining budget, and a final inference step resolves inconsistencies between the two sets of counts. 

In contrast to DP-compliant aggregate statistics published by a data curator, significant work has also been devoted to preventing the data curators themselves (such as a location based service provider) from inferring a mobile user's location in the online setting \cite{quercia2011spotme,ghinita2010reciprocal,andres2013geo}. Spatial $k$-anonymity (SKA)~\cite{ghinita2010reciprocal,gruteser2003anonymous} generalizes the specific position of the querying user to a region that encloses at least $k$ users. Geo-indistinguishability~\cite{andres2013geo} extends the DP definition to the Euclidean space and obfuscates user check-ins to protect the exact location coordinates. Synthesizing privacy-preserving location trajectories is explored in \cite{bindschaedler2016synthesizing}. Finally, reporting high-order statistics of mobility data in the context of DP has been pursued in ~\cite{Chen:KDD, Chen:CCS, zhang2016privtree}, where the focus is on releasing trajectories using noisy counts of prefixes or $n$-grams in a trajectory.

\vspace{-10pt}
\section{Conclusions and Future Work}
We proposed a novel approach to privacy-preserving release of location histograms with differential privacy guarantees. Our technique capitalizes on the key observation that density homogeneity within partitions of the constructed index structure reduces the error introduced by DP noise. We devised low-sensitivity strategies for finding split coordinates in a DP-compliant way, and implemented effective privacy budget allocation strategies across different stages of the data sanitization process. In future work, we plan to extend our approach to trajectory data, which are more challenging due to their high-dimensionality. We also plan to combine our data sanitization techniques with machine learning approaches, in order to further boost the accuracy of DP-compliant location queries.

\bibliographystyle{abbrv}
\bibliography{Ref_HVE}

\newpage
\newpage

\appendix
\section*{Appendix}

\renewcommand{\thesubsection}{\Alph{subsection}}

\subsection{Proof of Theorem 1}


Theorem {\em (Sensitivity of Partitioning)
    The sensitivity of cost function $o_k$ for any given horizontal or vertical index $k$ is $2$.
    }
\begin{proof}
In the calculation of objective function $o$ for a given index $k$, adding or removing an individual data point affects only one cell and the corresponding cluster. The objective function for split point $k$ can be written as
\begin{align}
    o_k = \sum_{i\Equal1}^k \sum_{j\Equal1}^V |c_{ij} - \mu_1| + \sum_{i\Equal k+1}^U \sum_{j\Equal1}^V |c_{ij} - \mu_2|,
\end{align}
The modified objective function value following addition of a single record to an arbitrary cell $c_{xy}$ can be represented as
\begin{align}
    {o_k^\prime} = \sum_{i\Equal1}^k \sum_{j\Equal1}^V |c_{ij}' - \mu_1'| + \sum_{i\Equal k+1}^U \sum_{j\Equal1}^V |c_{ij}'  - \mu_2'|.
\end{align}

Without loss of generality, assume that the additional record is located in the first cluster which results in $\mu_1' = \mu_1 + 1/kV$, $\mu_2' = \mu_2$, and $c_{ij}'$ being equal to $c_{ij}$ for all possible $i$ and $j$ except for $c_{xy}$ where we have $c_{xy}' = c_{xy} + 1$. Therefore, the sensitivity of the objective function has value 2 as follows:

\begin{align}\label{equ: delta}
    \Delta {o_k} = o_k - {o_k^\prime}  \leq \dfrac{2(kV-1)}{kV} \leq 2
\end{align}

\noindent Eq.~\eqref{equ: delta} is derived using the \textit{reverse triangle inequality}: 
\begin{equation}
\begin{split}
   &\Big{|} |c_{ij} - \mu_1 - \dfrac{1}{kV}|-   |c_{ij} - \mu_1| \Big{|} \leq \dfrac{1}{kV}\\ &\qquad \qquad \forall \{ ij | i\in\{1,...,k\} \land j\in \{1,...,V\}, ij\neq xy \}
\end{split}
\end{equation}
and
\begin{equation}
  \Big{|} | c_{xy} +1- \mu_1 - \dfrac{1}{kV}|-   |c_{xy} - \mu_1|  \Big{|}\leq \dfrac{kV-1}{kV} \qquad \quad
\end{equation}
Similarly, the sensitivity upper bound corresponding to an individual record's removal can be shown to be 2. 
\end{proof}
\balance

\end{document}